\documentclass[printer]{aa}
\usepackage{graphicx}
\usepackage{txfonts}
\def\gtrsim{\mathrel{\hbox{\rlap{\hbox{\lower4pt\hbox{$\sim$}}}\hbox{$>$}}}}
\newcommand{\mincir}{\raise -2.truept\hbox{\rlap{\hbox{$\sim$}}\raise5.truept
\hbox{$<$}\ }}
\newcommand{\magcir}{\raise -2.truept\hbox{\rlap{\hbox{$\sim$}}\raise5.truept
\hbox{$>$}\ }}
\newcommand{\siml}{\raise -2.truept\hbox{\rlap{\hbox{$\sim$}}\raise5.truept
\hbox{$<$}\ }}
\newcommand{\simg}{\raise -2.truept\hbox{\rlap{\hbox{$\sim$}}\raise5.truept
\hbox{$>$}\ }}
\newcommand{\be}{\begin{equation}}
\newcommand{\ee}{\end{equation}}
\newcommand{\ba}{\begin{eqnarray}}
\newcommand{\ea}{\end{eqnarray}}
\newcommand {\h} {$h^{-1}$ Mpc $ \;$}

\newcommand {\ks} {km~s$^{-1} \;$}
\newcommand {\kss} {km~s$^{-1}$}
\begin{document}
   \title{Galaxies in group and  field environments: \\ a comparison of
optical--NIR luminosities and colors}
%   \subtitle{}
   \author{M. Girardi\inst{1} 
\and F. Mardirossian\inst{1} 
\and C. Marinoni\inst{2} 
\and M. Mezzetti\inst{1} 
\and E. Rigoni\inst{1}}

   \offprints{M. Girardi}

   \institute{Dipartimento di Astronomia, Universit\`{a} degli Studi
di Trieste, Via Tiepolo 11, I-34131 Trieste, Italy\\
\email{girardi,mardiros,mezzetti,rigoni@ts.astro.it} \and Laboratoire
d'Astrophysique de Marseille, Marseille, France\\
\email{marinoni@astrsp-mrs.fr}} 
\date{} 
\abstract{ We compare properties of galaxies in loose groups with
those in field environment by analyzing the Nearby Optical Galaxy
(NOG) catalog of galaxy systems.  We consider as group galaxies,
objects belonging to systems with at least five members identified by
means of the ``friends of friends method'', and, as field galaxies,
all galaxies with no companions.  We analyze both a magnitude--limited
sample of 959 and 2035 galaxies (groups vs. field galaxies,
respectively, B$<14$ mag, and $2000<cz<6000$ \ks) and a
volume--limited sample ( $M_{\rm B} <-19.01+5\,{\rm log_{10}}\,{\rm
h}$ mag, $2000<cz<4000$ \ks 369 group and 548 field galaxies).  For
all these galaxies, blue corrected magnitudes and morphological types
are available.  The cross-correlation of NOG with the 2MASS second
release allow us to assign K magnitudes and obtain B--K colors for
about half of the galaxies in our samples.  We analyze luminosity and
color segregation--effects in relation with the morphological
segregation.  For both B and K bands, we find that group galaxies
are, on average, more luminous than field galaxies and this effect is
not entirely a consequence of the morphological
segregation.  After taking into account the morphological segregation,
the luminosity difference between group and field galaxies is about
$10\%$.  When considering only very early--type galaxies (T$<-2$) the
difference is larger than $30\%$.  We also find that group galaxies
are redder than field galaxies, $\Delta$(B--K) $\sim 0.4$ mag.
However, after taking into account the morphological segregation, we
find a smaller B--K difference, poorly significant (only at the c.l. of
$\sim 80\%$).  We discuss our results considering that the analyzed
groups define a very low density environment (projected mean density
$\sim 5-6$ h$^2$ Mpc$^{-2}$ galaxies).
}
\authorrunning{Girardi et al.} 
\titlerunning{Group vs. field
galaxies} \maketitle
%
%________________________________________________________________

\section{Introduction}
In spite of the variety and number of devoted studies, we are still
far from understanding how much of galaxy evolution is a reflection of
local physics on galactic scales, possibly in the context of biased
galaxy formation, and how much of it is due to environmental effects
intrinsically connected to the parent galaxy-system.

Most of past studies concerning the effects of the external
environment on galaxy evolution focused on central regions of galaxy
clusters and consistently showed that early--type galaxies (or very
luminous ones) inhabit preferentially cluster cores (e.g., Oemler
\cite{oem74}; Dressler \cite{dre80}; Biviano et al. \cite{biv92};
Dressler et al. \cite{dre97}; Biviano et al. \cite{biv02}). More
recent studies aim to analyze environmental effects as a function of
radius out to larger and larger distances from the cluster
center. These analyses probe the radial gradient in photometric and
spectroscopic properties out to or just beyond the virial radius: the
emerging picture is that galaxy star formation rate (SFR) is
suppressed via via galaxies are accreted onto clusters (e.g., Abraham
et al. \cite{abr96}; Balogh et al.  \cite{bal97}; Pimbblet et
al. \cite{pim02}).  Very recent analyses based on the 2dF Galaxy
Redshift Survey and on the Sloan Digital Sky Survey show that the
distribution of SFRs of cluster galaxies begins to change, compared
with the field population, at a clustercentric radius of 3-4 virial
radii (Lewis et al. \cite{lew02}; G\'omez et al. \cite{gom03}).
Moreover, the SFR of galaxies is strongly correlated with the local
projected galaxy--density down to a characteristic threshold (Lewis et
al. \cite{lew02}; G\'omez et al. \cite{gom03}).  These environmental
effects on the SFR are analogous to those on the galaxy morphology,
known as the morphology-radius and morphology-density relations (e.g.,
Oemler \cite{oem74}; Dressler \cite{dre80}; Postman \& Geller
\cite{pos84}; Whitmore et al. \cite{whi93}). Since SFR is lower in
early--type galaxies (e.g., Kennicutt \cite{ken83}; Jansen et
al. \cite{jan00}), which inhabit preferentially cluster cores and high
density regions, it is not clear whether the two effects are
completely independent.  First attempts suggest that morphology
segregation alone is unlikely to explain the SFR effect (Balogh et
al. \cite{bal98}; Lewis et al. \cite{lew02}; Pimbblet et
al. \cite{pim02}; G\'omez et al. \cite{gom03} ).

Both traditional and more recent approaches, concerning morphologies
and SFRs, respectively, consistently show that environmental
influences on galaxy properties are not restricted to cluster cores,
but are effective in groups, which could be the relevant sites of
galaxy evolution (e.g., Balogh \& Bower \cite{bal03}).  Although the
study of the group environment is a much more difficult task than the
study of cluster environment, there are now robust observational
evidences confirming that group galaxies of different morphological
type -- color -- spectral type are spatially segregated (e.g., Ozernoy
\& Reinhard \cite{oze76}; Postman \& Geller \cite{pos84}; Mahdavi et
al. \cite{mah99}; Tran et al. \cite{tra01}; Carlberg et
al. \cite{car01}; Dom\'{\i}nguez et al. \cite{dom02}).  Luminosity
segregation in groups is a more controversial issue (e.g., Ozernoy \&
Reinhard \cite{oze76}; Giuricin et al. \cite{giu82}; Mezzetti et
al. \cite{mez85}; Magtesyan \& Movsesyan \cite{mag95}).  Analyzing
loose groups identified in the Nearby Optical Galaxy (NOG) sample by
Giuricin et al. (\cite{giu01}), Girardi et al. (Girardi et
al. \cite{gir03}) have found evidence of morphology and B--band
luminosity segregation of galaxies both in space and in velocity, in
qualitative agreement with a continuum of segregation properties of
galaxy enbedded in systems, from low-mass groups to massive clusters.

The aim of this paper is to study the environmental dependence of
galaxy properties, luminosities and colors, in very low density
environments comparing loose groups and field in the NOG catalog.  In
particular, we use the large amount of morphological information
available for nearby galaxies to disentangle and discriminate in a
robust way luminosity and color segregation--effects from
morphological segregation.

The outline of this paper is as follows.  We describe the data sample
and magnitude data in Sects.~2 and 3, respectively.  We devote Sect.~4
to the analysis of luminosity and morphology segregation--effects, and
Sect.~5 to color segregation.  We discuss and summarize our results in
Sects.~6 and 7, respectively.

Unless otherwise stated, we give errors at the 68\% confidence level
(hereafter c.l.).

A Hubble constant of 100 $h$ \ks Mpc$^{-1}$ is used throughout.

\section{Group and field samples}

We analyze the NOG catalog (Giuricin et al. \cite{giu00}).  NOG is a
magnitude--limited catalog (corrected total blue apparent magnitude
B$\leq 14$), with an upper distance limit ($cz<6000$ \kss), which
contains $\sim 7000$ optical galaxies, basically extracted from the
Lyon--Meudon Extragalactic Database (LEDA; Paturel et
al. \cite{pat97}.  NOG covers about $2/3$ of the sky $(|b| >
20^{\circ})$, and is quasi-complete in redshift ($97\%$).  Almost all
NOG galaxies ($98.7\%$) have a morphological classification as taken
from LEDA, and parameterized by $T$ (the morphological--type code
system of RC3 catalog -- de Vaucouleurs et al. \cite{dev91}) with one
decimal figure.

Hereafter we consider only the galaxies with the full information
available: i) galaxy position; ii) radial velocity ${\rm v}=cz$, where
$z$ is the heliocentric redshift in the LG rest frame (according to
Yahil et al.  \cite{yah77}); iii) corrected total blue magnitude; iv)
morphology.

Group galaxies were identified by Giuricin et al. (\cite{giu00}) using
the ``friends--of--friends'' method.  In particular, by applying two
different variants of the percolation method, they obtained two
comparable catalogs of galaxy systems.  Here we use the P2 catalog
obtained by allowing both the distance and the velocity link
parameters to scale with distance (Huchra \& Geller \cite{huc82});
even if new 3D cluster finding algorithms have been recently proposed
(e.g., Marinoni et al. \cite{mar02}) this is the most frequently used
method of group identification for low redshift galaxies.

In order to obtain the best-quality group sample in our analysis we
remove: i) all groups identified as known clusters (see Table~7 of
Giuricin et al. \cite{giu00}); ii) all very poor groups with $n<5$
member galaxies. We remove clusters following our aim to study
low density environments.  Moreover, the clusters contained in the NOG
catalog are very few (only ten at $cz> 2000$ \ks), not particularly
rich (Abell richness $R\sim0$), and poorly sampled (10.5 is the median
number of members). Thus NOG clusters are not a very representative
sample of rich clusters to be useful in a comparison with group
results.  As for very poor groups, the efficiency of the percolation
algorithm has been repeatedly checked, showing that an appreciable
fraction of the poorer groups, those with $n<5$ members, might be
false (i.e. represent unbound density fluctuations), whereas richer
groups almost always correspond to real systems (e.g., Ramella et
al. \cite{ram89}; Ramella et al. \cite{ram95}; Mahdavi et
al. \cite{mah97}; Nolthenius et al. \cite{nol97}; Diaferio et
al. \cite{dia99}). The dilution effect of a significant number of
spurious groups in our analysis would hide or weaken any possible
difference between groups and field. For instance, this dilution
effect is suggested to explain some differences between groups with
$n<5$ and with $n\ge 5$ as concerning the segregation properties of
member galaxies (Girardi et al. \cite{gir03}, Sect.~4.3). In view of
this possible bias we prefer to remove $n<5$ groups from our sample.

We consider as field galaxies, all the NOG galaxies left unbound,
i.e. not belonging to any group or binary system.

Finally, we remove from our analysis all field galaxies with $cz\le
2000$ \ks and groups with $\overline{cz}\le 2000$ \ks, where the mean
group velocity $\overline{\rm{v}}=\overline{cz}$ is computed by using
the biweight estimator (Beers et al. \cite{bee90}).  In fact, where
the recession velocity is not dominant on its random component, it is
no longer a reliable indication of the distance.  We apply such a
conservative limit since our analysis requires accurate absolute
magnitudes. In particular, the value of $2000$ \ks is suggested by the
analysis of the velocity field in the local Universe where the effect
of peculiar velocities is higher for $cz<2000$ \ks (Marinoni et
al. \cite{mar98}). Moreover, thank to our lower distance limit, all
galaxies belonging to the main clumps and clouds of the Virgo cluster
are rejected, too (see Binggeli et al. \cite{bing87}; Binggeli et al.
\cite{bin93}; and Table~7 of Giuricin et al. \cite{giu00}).

Our final group sample contains 120 loose groups for a total of 959
galaxies (hereafter GROUP sample).  The field sample contains 2035
galaxies (FIELD sample). Table~\ref{tab1} lists the median values for
group main properties: number of members, $n$, and redshift, $z$.  We
also give the median values -- with corresponding $90\%$
confidence intervals -- and lower and upper quartiles of the
distributions for: group size, $R_{\rm{max}}$, which is the
(projected) distance of the most distant galaxy from the (biweight)
group center; LOS velocity dispersion, $\sigma_{\mathrm{v}}$, and
virial mass, ${\cal M}$, computed following Girardi \& Giuricin
(\cite{gir00}; see also Girardi et al. \cite{gir03}).  The
confidence intervals are computed following the procedure\footnote{
For the median of an ordered distribution of $N$ values the confidence
intervals $x_{(r)}$ and $x_{(N-r+1)}$, corresponding to a probability
$P(x_{(r)}\le x\le x_{(N-r+1)})=1-\alpha$, can be obtained from
$1-\alpha=2^{-N} \sum^{N-r}_{i=r} \left({N}\atop {i}\right)$.}
described by Kendall \& Stuart (\cite{ken79}, eq. 32.23) and first
proposed by Thompson (\cite{tho36}).

 \begin{table*}
      \caption[]{Group properties}
         \label{tab1}

%%%%%%%%%%%%%%%%%%%%%%%%%%%%%%%%%%%%%%%%%%%%%%%%%%%%%%%%%%%%%%%%%%%%%%%
%
%                        TABLE 1
%
%%%%%%%%%%%%%%%%%%%%%%%%%%%%%%%%%%%%%%%%%%%%%%%%%%%%%%%%%%%%%%%%%%%%%%%

\begin{tabular}{cccccccccc} 
\hline \hline
\multicolumn{1}{c}{$N_{\rm{GROUPs}}$} 
&\multicolumn{1}{c}{$N_{\rm{GALs}}$} 
&\multicolumn{1}{c}{$n$}
&\multicolumn{1}{c}{$z$}
&\multicolumn{2}{c}{$R_{\rm{max}}$}
&\multicolumn{2}{c}{$\sigma_{\rm{v}}$}
&\multicolumn{2}{c}{$\cal M$}
\\
\multicolumn{1}{c}{} 
&\multicolumn{1}{c}{} 
&\multicolumn{1}{c}{} 
&\multicolumn{1}{c}{}
&\multicolumn{2}{c}{\h}
&\multicolumn{2}{c}{km s$^-1$}
&\multicolumn{2}{c}{$h^{-1}\,10^{13}\,\cal M_{\sun}$}
\\
\multicolumn{1}{c}{} 
&\multicolumn{1}{c}{} 
&\multicolumn{1}{c}{}
&\multicolumn{1}{c}{}
&\multicolumn{1}{c}{Median}
&\multicolumn{1}{c}{Lower, Upper}
&\multicolumn{1}{c}{Median}
&\multicolumn{1}{c}{Lower, Upper}
&\multicolumn{1}{c}{Median}
&\multicolumn{1}{c}{Lower, Upper}
\\
\hline
120&959&6&0.013&$0.70^{+0.05}_{-0.08}$&0.54, 0.90&$179^{+19}_{-21}$&119, 260&$2.2^{+0.5}_{-0.6}$&0.7, 5.2\\
\hline
\end{tabular}

   \end{table*}

Moreover, we also extract from the above magnitude--limited sample a
volume--limited subsample, which by definition contain objects that
are luminous enough to be included in the sample when placed at the
cutoff distance.  We define the volume--limited samples of group and
field galaxies with depth of 4000 \ks by considering only galaxies
with blue corrected absolute--magnitude $M_{\rm B}<-19.01$ (hereafter
VLGROUP and VLFIELD samples, respectively).  The limit of 4000 \ks is
again suggested by the analysis of Marinoni et al. (\cite{mar98}),
since the discrepancy between different peculiar velocity models
derived using independent techniques is more pronounced for $cz>4000$
\ks.  The VLGROUP and VLFIELD ($2000 < cz < 4000$\ks) samples contain
369 and 548 galaxies, respectively.

\begin{figure} 
\centering   
\resizebox{\hsize}{!}{\includegraphics{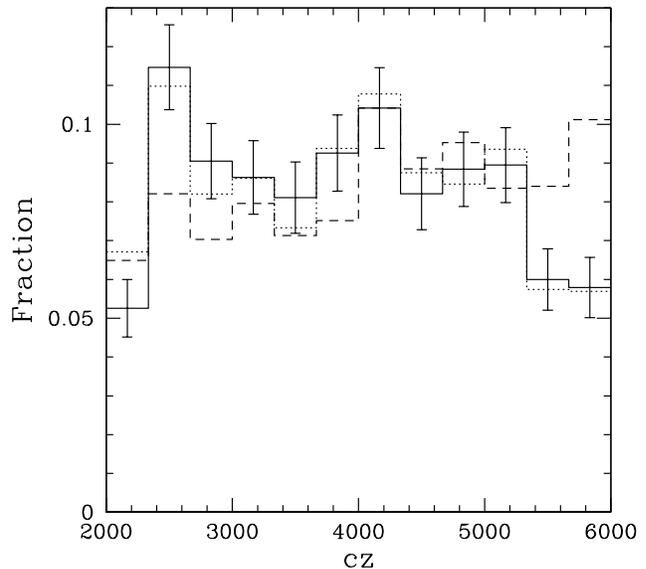}}
\caption{Comparison of $cz$ distributions
of group and field galaxies (solid and dashed lines, respectively).
The error bars are 1sigma Poissonian errors.
The dotted line indicates the simulated field (S-FIELD, see text)}.
\label{fig1}
\end{figure}

As for the comparison of GROUP vs.  FIELD samples we have to check for
possible biases connected with depth since these samples are extracted
from an (apparent-) magnitude--limited sample (see Sect.~4.2, too).
Fig.~\ref{fig1} shows the comparison of radial velocity distributions
for field and group galaxies. Small differences are shown, probably
due to very local differences in the clustering properties of large
scale structure. Analyzing the corresponding cumulative distributions
we find a difference at the 99.99\% c.l., according to the
Kolmogorov--Smirnov test (hereafter KS-test; e.g., Ledermann
\cite{led82}). To avoid possible observational biases we construct a
simulated sample of field galaxies which mimics the velocity
distribution of group galaxies. We use the inverse
transformation method for generating a random deviate $y$ from a known
probability distribution $f(y)$ as outlined in, e.g., Press et
al. (\cite{pre92}).  In practice, given the (normalized) distribution
of velocities of group galaxies (in the range 0-6000 \ks), $f({\rm
v})$, and the cumulative distribution, $F({\rm v})=\int_{0}^{\rm
v}f({\rm v})d{\rm v}$, we choose an uniform random $x=F({\rm v})$: its
corresponding ${\rm v}$ is the desired deviate and we assign to the
simulated field a galaxy of the real field very close to that velocity
${\rm v}$, i.e. one galaxy randomly selected in a range of $20$ \ks.
In this way we assign to the simulated field ten thousands of
galaxies, of which $\sim 6000$ are those with $cz> 2000$ \ks.  As
expected, there is no significant difference between the velocity
distribution of simulated--field galaxies and that of group galaxies
(see Fig.~\ref{fig1}).  Although general results are very similar
(see Sect.~4.2), hereafter we consider the simulated field, S-FIELD,
in any comparison with GROUP sample, after properly rescaling the
number of simulated objects to that of real--field galaxies when
statistical confidence levels are derived.

Note that in the above resampling we have used velocities of
individual group galaxies rather than luminosity distances, the latter
being based on mean velocities of the corresponding parent group.  Our
choice is due to the requirement of using a variable with a smooth,
well-behaved probability distribution. In fact, the probability
distribution of luminosity distances for group galaxies shows many
peaks, which are higher for groups with a larger number of members.
From the physical point of view, we use the KS-test to verify that the
cumulative distribution of individual velocities of group galaxies
does not significantly differ from that obtained substituting
individual velocities with mean velocities. Thus any possible effect
due to the chosen variable should be negligible.

\section{Galaxy magnitudes and colors}

NOG total, corrected B--band magnitudes come from LEDA compilation,
which collects and homogenize data of several catalogs.  They have
been converted to the standard systems of the RC3 catalog (de
Vaucouleurs et al. \cite{dev91}) and have been corrected for Galactic
extinction, internal extinction, and $k$--dimming (see Paturel et
al. \cite{pat97} for more details).

We take K$_s$-band survey magnitudes, K$_{20}$, from the extended
catalog of the official 2MASS second release. To assign the
counterpart of each galaxy in our sample we choose the closest 2MASS
object within a search radius of 0.3 arcmin. We use this search radius
to enclose also very nearby, with large angular size, galaxies. The
typical angular separation between a NOG galaxy and its counterpart is
much smaller ($\sim 4$-5 arcsecs).  We find counterparts for 3263 of
7076 galaxies of the whole NOG catalog. Out of these 3263 2MASS
objects, the possible contamination of non-galaxy objects is quite
small: only 56 objects have parameters which indicate a possible
artifact or contaminated and/or confused source, i.e.  $e_{\rm
score}<1.4$, $g_{\rm score}<1.4$, $cc_{\rm flag}\ne0$ (see Jarrett et
al. \cite{jar00}, Cole et al. \cite{col01}).

\begin{figure} 
\centering  
\resizebox{\hsize}{!}{\includegraphics{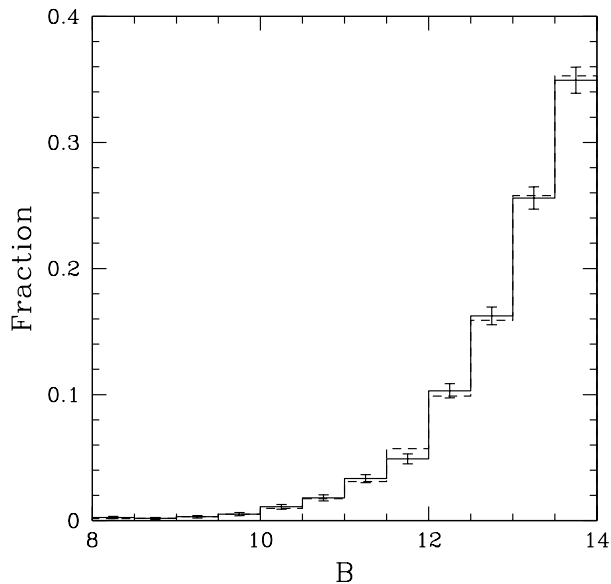}}
\caption{Comparison between magnitude distribution
of all NOG galaxies with that of matched 2MASS-NOG galaxies
(solid and dashed lines, respectively).
The error bars are 1sigma Poissonian errors.}
\label{fig2}
\end{figure}

The fraction of matched galaxies corresponds to the incomplete sky
coverage in the 2MASS 2nd release ($\sim 47\%$ of the sky).  Moreover,
the B-magnitude distribution of matched 2MASS-NOG galaxies is not
different from that of all NOG galaxies suggesting that 2MASS-NOG
galaxies form an unbiased subsample of the whole NOG catalog
(see Fig.~\ref{fig2}).

As for our specific samples, we obtain K$_{20}$ magnitude for 440 and
936 galaxies in GROUP and FIELD samples, respectively (167 and 240
galaxies in VLGROUP and VLFIELD, respectively).

To obtain total corrected K magnitudes we consider the offset of 0.2
mag between isophotal and total magnitudes (see Kochanek et
al. \cite{koc01}) and we correct for Galactic extinction, internal
extinction and $k$--dimming:
\begin{equation}
{\rm K}={\rm K}_{20}-0.2-R_{\rm K}{\rm [E(B}{\rm -}{\rm V)}+A_{\rm i,B}/4]-k(z).
\end{equation}
\noindent 
We correct for Galactic extinction, $A_{\rm K}=R_{\rm K}$E(B-V)],
using the color excess E(B-V) recovered by the extinction maps of
Schlegel et al.  (\cite{sch98}) and an extinction coefficient of
$R_{\rm K}=0.35$ (Cardelli et al. \cite{car89}).  For the term of
internal extinction $A_{\rm i}=R_{\rm K }[{\rm E}_{\rm i} {\rm(B{\rm
-V})}]$ we assume E$_{\rm i}$(B--V)$\sim A_{\rm i,B}/4$, where the
internal B--band absorption of galaxies is $A_{\rm i,B}=0$ for $T<0$
and $A_{\rm i,B}=0.4$ for $T>0$ according to the mean values suggested
by RC3 for different morphological types.  Finally, we apply the
K$_s$-band $k$--correction of $k(z)=-6.0$\,log$(1+z)$, independent of
galaxy type and valid for $z\lesssim 0.25$ (based on the Worthey
\cite{wor94} models, see Kochanek et al. \cite{koc01}).

We convert B-- and K--band magnitudes to absolute magnitudes $M_{\rm B}$
and $M_{\rm K}$ by using the luminosity distance recovered from redshift
$z$.  For group galaxies we use the $\overline{cz}$ of the parent group.

\section{Luminosity and morphology environmental effects}

\subsection{Analysis and results}

\begin{figure} 
\centering 
\resizebox{\hsize}{!}{\includegraphics{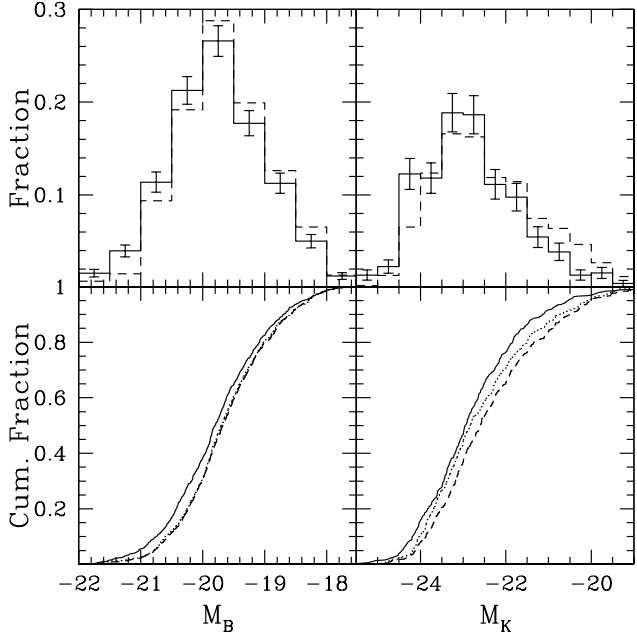}}
\caption{Comparison between absolute magnitude distributions of group
and field galaxies in B and K bands (left and right panels,
respectively) for GROUP and S-FIELD samples (solid and dashed lines,
respectively). The error bars in the top-panels are 1sigma
Poissonian errors.  The corresponding cumulative distributions are
shown in the bottom panels: there, the dotted lines indicate the
distributions for the simulated field ST-S-FIELD, which mimics
morphological--type distribution of group galaxies (see text).  }
\label{fig3}
\end{figure}

\begin{figure} 
\centering 
\resizebox{\hsize}{!}{\includegraphics{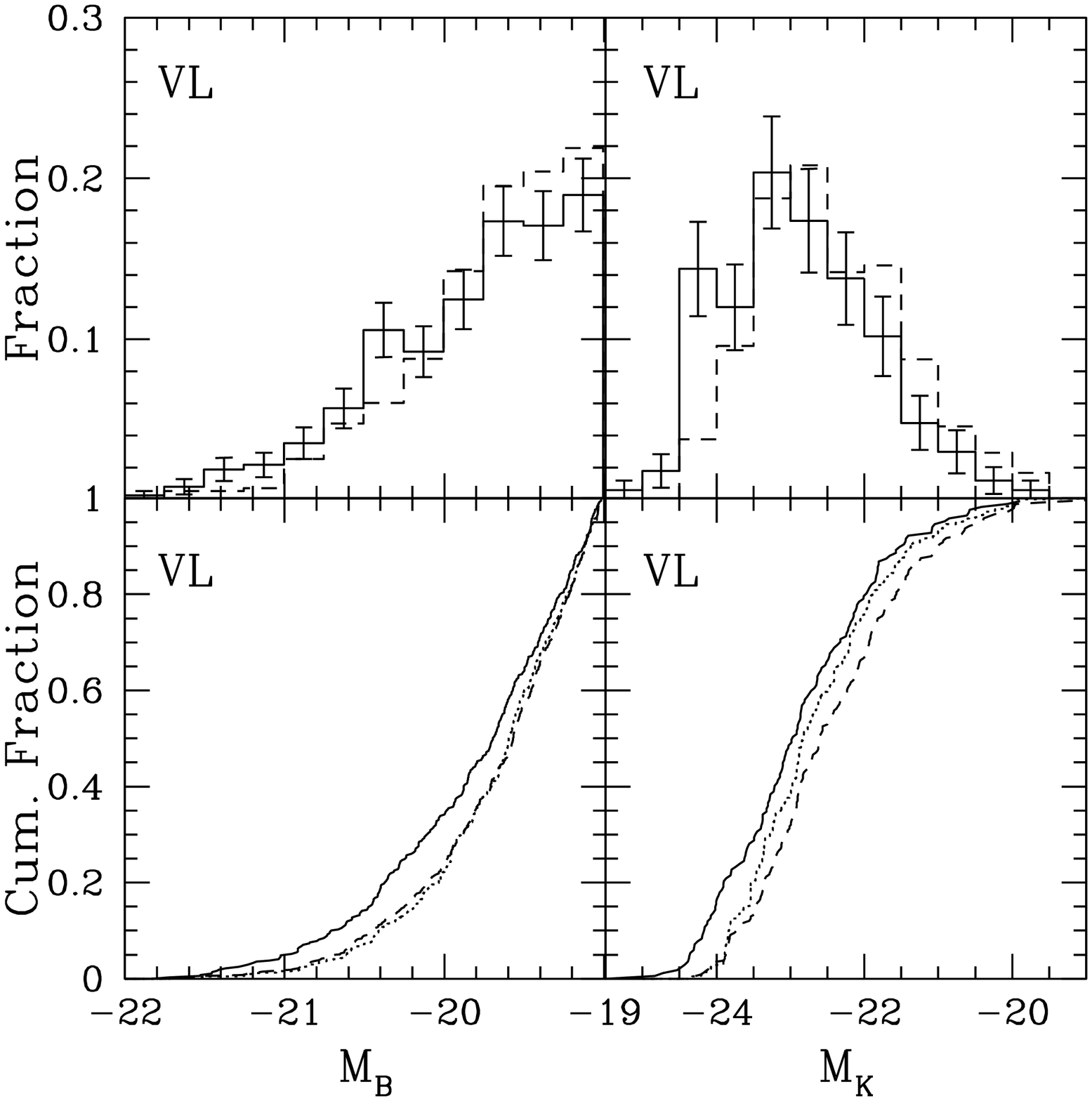}}
\caption{Comparison between absolute magnitude distributions of group
and field galaxies in B and K bands (left and right panels,
respectively) for VLGROUP and VLFIELD samples (solid and dashed lines,
respectively). The error bars in the top-panels are 1sigma
Poissonian errors.  The respective cumulative distributions are shown
in the bottom panels: there, the dotted lines indicate the
distributions for the simulated field ST-VLFIELD, which mimics
morphological--type distribution of group galaxies (see text).}
\label{fig4}
\end{figure}

Figs.~\ref{fig3} and \ref{fig4} show that group galaxies are more
luminous in both B and K bands than field galaxies.  To compute the
significance of this difference, we compare the cumulative
distributions of B magnitudes for GROUP and S-FIELD (VLGROUP
and VLFIELD): they are different at the $>99.99\%$ ($99.62\%$)
according to the KS-test; the same result is found for K magnitudes.

\begin{figure} 
\centering 
\resizebox{\hsize}{!}{\includegraphics{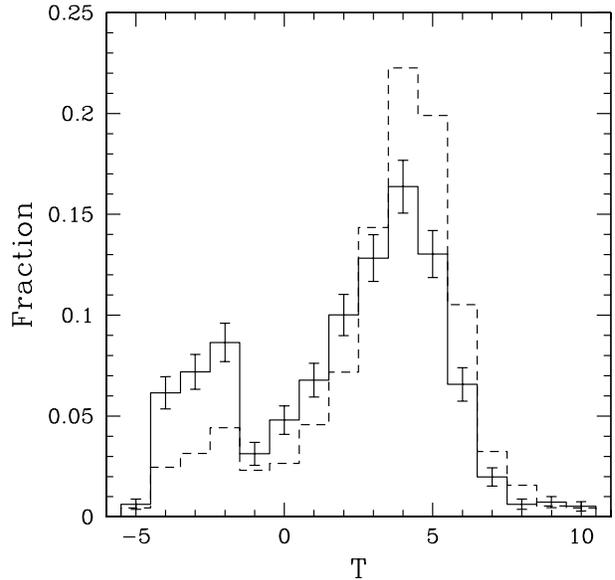}}
\caption{Comparison between morphological--type distribution
of GROUP and FIELD galaxies 
(solid and dashed lines, respectively).
The error bars are 1sigma Poissonian errors.}
\label{fig5}
\end{figure}

The morphological distributions in group and field environments are
different at the $>99.99\%$ c.l. (according to the KS-test).  In fact,
the fraction of early galaxies in groups is larger than in the field
(see Fig.~\ref{fig5}).  In particular, we compute the relative
proportion of E+S0 ($T<-2$), S0+S$_{\rm{early}}$ ($-2\le T<1$),
S$_{\rm{middle}}$ ($1\le T<4$), and S$_{\rm{late}}$+I ($T\ge 4$)
obtaining (18:16:35:31) and (8:8:35:49) for GROUP and FIELD
environments, respectively. The morphological--type distribution of
S-FIELD galaxies is not different from that of FIELD.

\begin{figure} 
\centering  
\resizebox{\hsize}{!}{\includegraphics{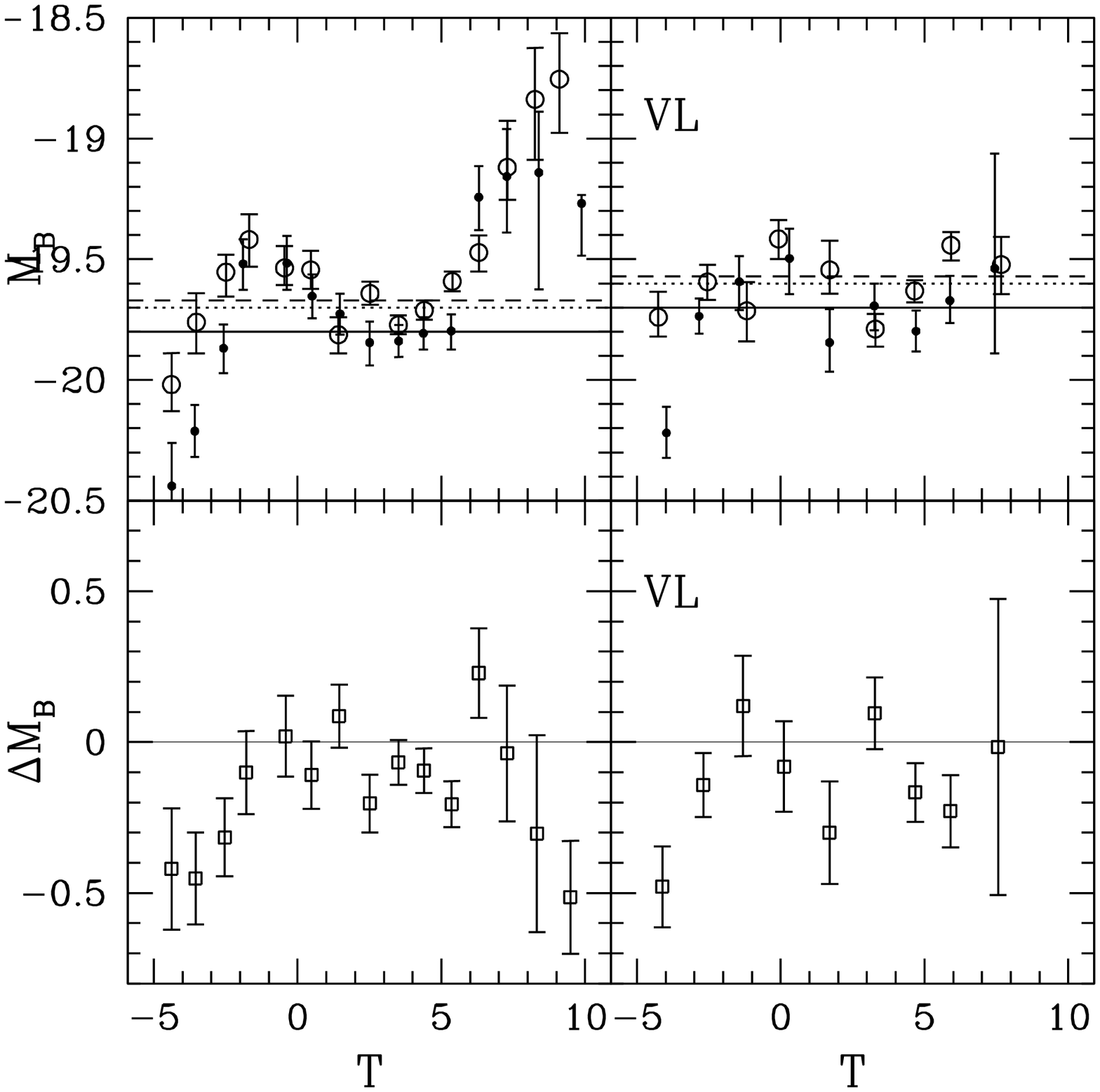}}
\caption{Top panels: B--band absolute--magnitude as a function of
morphological type for GROUP and S-FIELD (left panels) and VLGROUP and
VLFIELD (right panels).  Points are biweight mean values for galaxies
in groups (filled circles) and field (open circles).  Error bars are
$68\%$ bootstrap estimates.  Only bins containing more than three
galaxies both for group and field samples are plotted.  Solid,
dashed, and dotted lines indicate the median values of magnitude
distributions of group, field, and simulated field galaxies,
respectively; see also left--down panels of Figs.~\ref{fig3} and
\ref{fig4}.  Bottom panels show the magnitude difference between
group and field as recovered from the top panels.}
\label{fig6}
\end{figure}

\begin{figure} 
\centering  
\resizebox{\hsize}{!}{\includegraphics{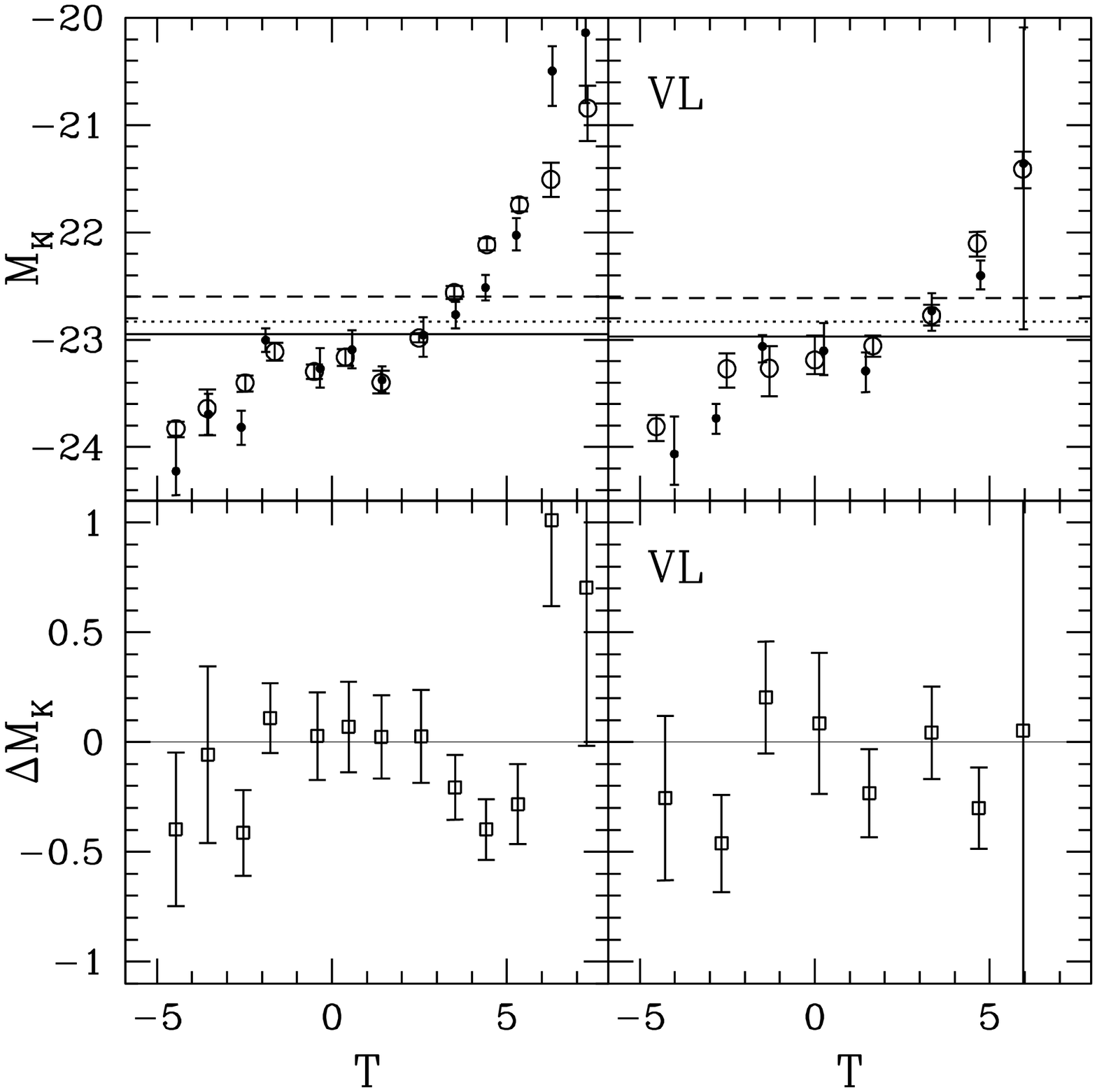}}
\caption{Top panels: K--band absolute--magnitude as a function of
morphological type for GROUP and S-FIELD (left panels) and VLGROUP and
VLFIELD (right panels).  Points are biweight mean values for galaxies
in groups (filled circles) and field (open circles).  Error bars are
$68\%$ bootstrap estimates. Only bins containing more than three
galaxies both for group and field samples are plotted.  Solid,
dashed, and dotted lines indicate the median values of magnitude
distributions of group, field, and simulated field galaxies,
respectively; see also right--down panels of Figs.~\ref{fig3} and
\ref{fig4}.  Bottom panels show the magnitude difference between
group and field as recovered from the top panels.  }
\label{fig7}
\end{figure}

To disentangle luminosity and morphological segregation--effects we
consider mean luminosity values as a function of morphological types,
see Figs.~\ref{fig6} and \ref{fig7} (top panels). The difference is
better outlined in the corresponding bottom panels where $\Delta
M=M_{\rm{groups}}-M_{\rm{field}}$ is shown: under the same
morphological type, group galaxies are more luminous than the
respective field galaxies. According to the $\chi^2$-test, the null
hypothesis $\Delta M_{\rm B}=0$ is rejected at the $>99.99\%$
($99.78\%$) when comparing the GROUP and S-FIELD samples (VLGROUP and
VLFIELD).  As for K magnitudes, the null hypothesis $\Delta M_{\rm
K}=0$ is rejected at the $98.61\%$ ($70.34\%$) when comparing the
GROUP and S-FIELD samples (VLGROUP and VLFIELD).

We prefer to avoid the arbitrary binning choice intrinsically
connected to the use of the $\chi^2$-test. Consequently we apply an
alternative statistical method.  We construct a simulated field (of
10000 galaxies) in order to mimic the morphological distribution of
group galaxies and we do this by using the same technique outlined in
Sect.~2. That is, after having selected a random deviate $T$ from the
whole morphological--type distribution of GROUP (VLGROUP) galaxies, we
assign to the simulated field a galaxy of the S-FIELD (VLFIELD) very
close to that $T$, i.e. one galaxy randomly selected in a range of
$\Delta T=0.25$.  In this way we construct a simulated ST-S-FIELD
(ST-VLFIELD) sample which has a $T$-distribution similar to that of
the GROUP (VLGROUP) sample. Figs.~\ref{fig3} and \ref{fig4} (bottom
panels) show the cumulative magnitude distribution of this simulated
field for the magnitude-- and volume--limited samples,
respectively. According to the KS-test, we find a difference of
$99.98\%$ ($99.94\%$) when comparing B--magnitude distributions of
GROUP and ST-S-FIELD (VLGROUP and ST-VLFIELD) and of $94.55\%$
($99.40\%$) when comparing K--magnitude distributions of GROUP and
ST-S-FIELD (VLGROUP and ST-VLFIELD). The effect of resampling the
field galaxies as described above is different in B and K band, being
much smaller in the first case.  The large effect in the K band is due
to the fact that the galaxies which inhabit preferentially field
environment (i.e., $T\gtrsim3$ galaxies, see Fig.~\ref{fig5}) are
generally less luminous than the median value of the field magnitude
distribution (see dashed lines in the top panels of Fig.~\ref{fig7}),
while the galaxies which are rare in field environment ($T\lesssim3$)
are more luminous than the median field magnitude.  Thus both these
populations contribute to raise the global K--band luminosity when
their fraction is renormalized to that of group environment in the
resampling procedure.  As for the B band, the two populations of
$3\lesssim T\lesssim 5$ and $-2\lesssim T\lesssim 0$ galaxies, which
are more and less luminous than the median magnitude, respectively
(see Fig.~\ref{fig6}, top panels), lower the global luminosity in the
resampling procedure, thus counterbalancing the effect of other galaxy
populations.

The above simulated fields allows us to give an estimate of the
typical amount of magnitude difference between group and field
galaxies which is independent of the morphological segregation.  This
amount can be obtained from the difference of the median values of
magnitude distributions: $\Delta M_{\rm B}=-0.10$ mag (for GROUP
vs. ST-S-FIELD), and $\Delta M_{\rm K}=-$[0.12---0.14] mag (for
VLGROUP and ST-VLFIELD), see Figs.~\ref{fig6} and ~\ref{fig7} (top
panels).

Figs.~\ref{fig6} and \ref{fig7} suggest that the above difference is
mainly connected to early--type galaxies.  We compute the mean
magnitude values and respective errors obtained for each morphological
class for both the magnitude-- and volume--limited samples.
Table~\ref{tab2} lists: the sample name; the number of galaxies having
B magnitude, $N_{\rm{GALs,B}}$:g,f (in groups and in the field); the
mean B absolute--magnitude and corresponding error for group and
galaxies, $M_{\rm{B,group}}$ and $M_{\rm{B,field}}$, respectively; the
number of galaxies having K magnitude, $N_{\rm{GALs,K}}$:g,f (in
groups and in the field); the mean K absolute--magnitude and
corresponding error for group and galaxies, $M_{\rm{K,group}}$ and
$M_{\rm{K,field}}$, respectively; the mean B--K color and
corresponding error for group and galaxies, (B--K)$_{\rm group}$ and
(B--K)$_{\rm field}$, respectively.

The only 3sigma difference we find between group and field values is
for E+S0 galaxies in the B band: $\Delta M_{\rm B}=-0.36\pm 0.08$ and
$-0.32\pm 0.09$ mag for the magnitude-- and volume--limited samples,
respectively.  As far as the K band is concerned, the amount of the
difference is similar, but significant at the 2.4sigma c.l..

 \begin{table*}
      \caption[]{Absolute magnitudes for different morphological types}
\label{tab2}
%%%%%%%%%%%%%%%%%%%%%%%%%%%%%%%%%%%%%%%%%%%%%%%%%%%%%%%%%%%%%%%%%%%%%%%
%
%                        TABLE 1
%
%%%%%%%%%%%%%%%%%%%%%%%%%%%%%%%%%%%%%%%%%%%%%%%%%%%%%%%%%%%%%%%%%%%%%%%

\begin{tabular}{lcccccccc} 
\hline \hline
\multicolumn{1}{c}{Sample} 
&\multicolumn{1}{c}{$N_{\rm{GALs,B}}$:g,f} 
&\multicolumn{1}{c}{$M_{\rm{B,group}}$}
&\multicolumn{1}{c}{$M_{\rm{B,field}}$}
&\multicolumn{1}{c}{$N_{\rm{GALs,K}}$:g,f} 
&\multicolumn{1}{c}{$M_{\rm{K,group}}$}
&\multicolumn{1}{c}{$M_{\rm{K,field}}$}
&\multicolumn{1}{c}{(B--K)$_{\rm group}$}
&\multicolumn{1}{c}{(B--K)$_{\rm field}$}
\\
\multicolumn{1}{c}{} 
&\multicolumn{1}{c}{} 
&\multicolumn{1}{c}{} 
&\multicolumn{1}{c}{} 
&\multicolumn{1}{c}{}
&\multicolumn{1}{c}{} 
&\multicolumn{1}{c}{} 
&\multicolumn{1}{c}{}
&\multicolumn{1}{c}{}
\\
\hline
mag.lim. E+S0               &170,172&$-20.03\pm0.06$&$-19.67\pm0.06$& 72,73 &$-23.78\pm0.11$&$-23.39\pm0.11$&$3.88\pm0.04$&$3.73\pm0.06$\\
mag.lim. S0+S$_{\rm{early}}$&151,172&$-19.56\pm0.06$&$-19.48\pm0.05$&71,86  &$-23.16\pm0.08$&$-23.18\pm0.08$&$3.74\pm0.05$&$3.83\pm0.04$\\
mag.lim. S$_{\rm{middle}}  $&339,703&$-19.80\pm0.04$&$-19.73\pm0.03$&162,345&$-22.92\pm0.08$&$-22.64\pm0.06$&$3.06\pm0.05$&$2.93\pm0.04$\\
mag.lim. S$_{\rm{late}}$+I  &299,988&$-19.68\pm0.05$&$-19.57\pm0.02$&135,432&$-21.91\pm0.11$&$-21.81\pm0.06$&$2.23\pm0.08$&$2.16\pm0.04$\\
VL    E+S0               &69,53  &$-19.98\pm0.07$&$-19.66\pm0.05$&32,19  &$-23.77\pm0.11$&$-23.39\pm0.13$&$3.89\pm0.06$&$3.74\pm0.04$\\
VL    S0+S$_{\rm{early}}$&43,39  &$-19.67\pm0.08$&$-19.58\pm0.07$&20,17  &$-23.24\pm0.17$&$-23.21\pm0.11$&$3.65\pm0.09$&$3.77\pm0.03$\\
VL    S$_{\rm{middle}}  $&129,187&$-19.85\pm0.06$&$-19.80\pm0.04$&59,97  &$-22.93\pm0.12$&$-22.79\pm0.08$&$3.11\pm0.08$&$3.02\pm0.03$\\
VL    S$_{\rm{late}}$+I  &128,269&$-19.80\pm0.05$&$-19.64\pm0.03$&56,107 &$-22.22\pm0.13$&$-21.84\pm0.09$&$2.36\pm0.12$&$2.19\pm0.04$\\
\hline
\end{tabular}
   \end{table*}

\subsection{About luminosities comparison}

Since our above analysis requires the computation of absolute
magnitudes, it is worth to verify the robustness of our results in
relation to the distance estimates involved.

\begin{figure} 
\centering  
\resizebox{\hsize}{!}{\includegraphics{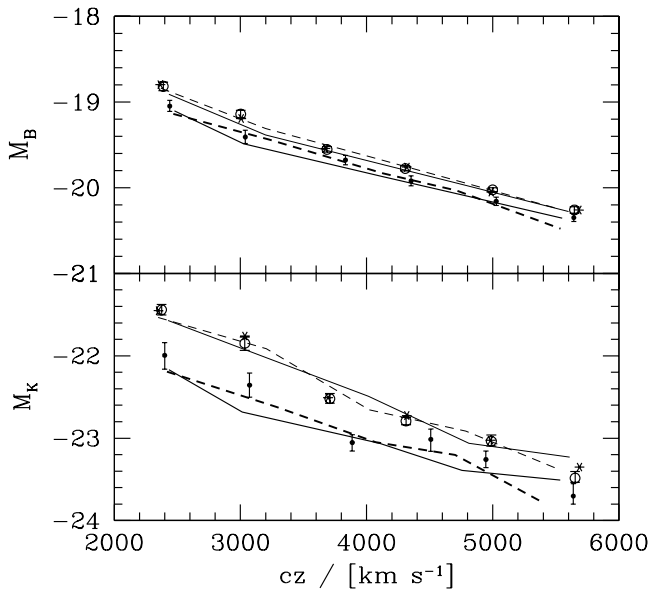}}
\caption{Absolute magnitude as function of $cz$ for B and K bands (top
and bottom panels, respectively).  Points are biweight mean values for
galaxies in GROUP sample (filled circles) and S-FIELD sample (open
circles).  Error bars are 68\% bootstrap estimates.  The results for
the real (not simulated) field are also shown (stars).
Solid and dashed lines show the trend for the two alternative $cz$ estimates,
recovered by using two different corrections for the peculiar velocity field
(corrections 1 and 2, respectively, see text)
for group and field galaxies (thick and thin lines, respectively).
}
\label{fig8}
\end{figure}

The question of the analysis in our largest sample, i.e. the
(apparent-) magnitude--limited one, is the most complex.  In fact, by
definition of magnitude--limited samples, only intrinsically more
luminous galaxies are sampled in more distant volumes and thus a
particular result concerning galaxy luminosities should be always
verified in order to avoid any spurious dependence on the catalog
depth.  Fig.~\ref{fig8} shows the behavior of absolute magnitudes as
function of $cz$: group galaxies are systematically more luminous than
field galaxies.  However, to avoid any possible bias we use throughout
our analysis the simulated field S-FIELD, which mimics the $cz$
distribution of group galaxies (see Sect.~2).  Fig.~\ref{fig8} shows
that the real and simulated fields are effectively indistinguishable;
thus our use of S-FIELD should be considered a very prudential
approach, which does not bias our main conclusions.

A more general question concerns the estimate of the galaxy distances.
We verify the robustness of our results by recomputing absolute
magnitudes on the basis of two alternative distance estimates derived
using models of the peculiar velocity field in the local Universe
(Marinoni et al. \cite{mar98}).  Marinoni et al. used two independent
approaches to model the peculiar velocity field and correct the
redshift--dependent distances for peculiar motions: (1) a
multiattractor model fitted to the Mark III catalog of peculiar
velocities (Willick et al. \cite{wil95}; \cite{wil96}; \cite{wil97});
(2) a cluster dipole reconstruction scheme (Branchini \& Plionis
\cite{bra96}) modified with the inclusion of a local model of
Virgocentric infall.  We have used these models in order to transform
the observed redshift of NOG galaxies into "pseudo-real" distances.
Fig.~\ref{fig8} shows the resulting behavior of absolute magnitudes as
a function of the peculiar-velocity corrected redshifts ($cz$): the
trend is consistent with the one inferred using the original
uncorrected $cz$ in our analysis; therefore we conclude that the
details of the adopted distance estimate have not biased our results
in any noticeable way.

\section{B--K color and environment}

\begin{figure} 
\centering 
\resizebox{\hsize}{!}{\includegraphics{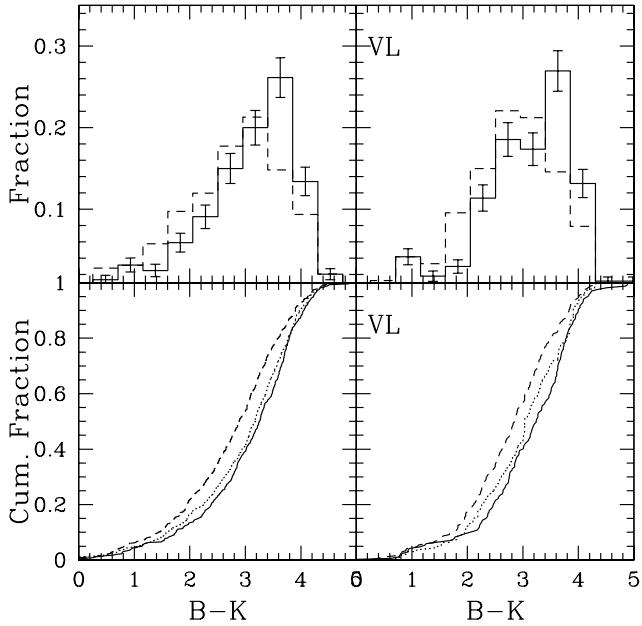}}
\caption{Comparison between B--K distributions of group and field
galaxies (solid and dashed lines, respectively) for GROUP and S-FIELD
(left panels) and VLGROUP and VLFIELD (right panels).  The error bars
in the top-panels are 1sigma Poissonian errors.  The corresponding
cumulative distributions are shown in the bottom panels: there, the dotted
lines indicate distributions of the simulated fields, ST-S-FIELD and
ST-VLFIELD, which mimic morphological--type distribution of group
galaxies (see text).}
\label{fig9}
\end{figure}

\begin{figure} 
\centering  
\resizebox{\hsize}{!}{\includegraphics{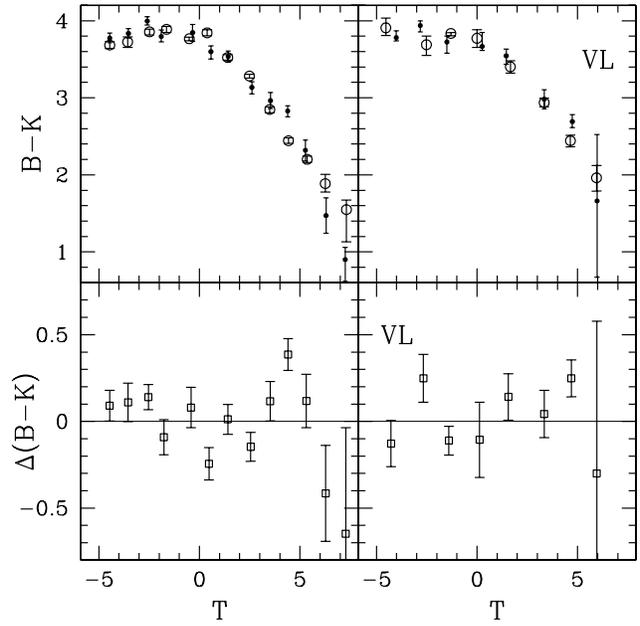}}
\caption{Top panels: B--K color as a function of morphological
type for GROUP and S-FIELD (left panels) and VLGROUP and VLFIELD
(right panels).  Points are biweight mean values for galaxies in
groups (filled circles) and field (open circles).  Error bars are
$68\%$ bootstrap estimates.  Only bins containing more than three
galaxies both for group and field samples are plotted.  Bottom panels
show the color difference between group and field as recovered from
the top panels.}
\label{fig10}
\end{figure}

Fig.~\ref{fig9} compares B--K colors of galaxies in group and field
environment.  Group galaxies have larger B--K colors than field
galaxies: the cumulative distributions are different at the $>99.99\%$
and $>99.95\%$ c.l. according to the KS-test for the magnitude-- and
volume--limited samples, respectively.  The amount of color difference
between groups and field is $\Delta$(B--K)=(B--K)$_{\rm
groups}$--(B--K)$_{\rm field} \sim 0.4$ mag.  Fig.~\ref{fig10} shows
the mean B--K values as a function of morphological type in order to
disentangle color and morphology segregation-effects.

As in Sect.~4.1, to obtain a quantitative result, we compare the
cumulative distributions of galaxy colors in groups and in the
simulated-field which mimics the morphological distribution of group
galaxies (ST-S-FIELD or ST-VLFIELD), see  bottom panels in
Fig.~\ref{fig9}.  The amount of the B--K difference between groups and
the simulated field is much smaller, $\Delta$(B--K)=0.07-$0.21$ mag
for the magnitude-- and volume--limited samples, respectively.
According to the KS-test, the statistical significance of the
difference is very low: $84.58\%$ and $76.98\%$ for GROUP
vs. ST-S-FIELD, and VLGROUP vs. ST-VLFIELD, respectively.

\section{Discussion and conclusions}

Different authors (e.g., Marinoni et al. \cite{mar99}; Ramella et
al. \cite{ram99}) have analyzed the luminosity function of galaxies in
low and high density environments concluding that group galaxies are,
on average, more luminous than field galaxies.  By using a
complementary approach, we confirm this environmental dependence and
show that this effect is  not entirely a consequence of
the morphological segregation.

This result is an extension at lower-density environments of the
results obtained for the internal regions of galaxy systems, where
more luminous galaxies lie preferentially in more central regions,
independently of morphological segregation. Luminosity and morphology
segregations are well supported for clusters (e.g., Adami et
al. \cite{ada98}; Biviano et al. \cite{biv02}) and similar results are
found in NOG groups (Girardi et al. \cite{gir03} and refs. therein).
Within cluster environment, luminosity segregation seems to concern
only ellipticals and possibly lenticulars (e.g. Biviano et
al. \cite{biv92}; Stein \cite{ste97}; Biviano et al. \cite{biv02}),
but a definitive conclusion in not reached in NOG groups.  From the
present study (see Figs.~\ref{fig6} and \ref{fig7}) we obtain that
luminosity segregation, in particular in the B band, is particularly
strong for very early--type galaxies ($T\lesssim -2$).  When
considering only these early--type galaxies the luminosity difference
between group and field galaxies is larger than $30\%$.  The global
difference, taking into account the morphological segregation, is much
more modest, about $10\%$, but still statistically very significant.

Since a large fraction of galaxies is embedded in groups ($\sim 40\%$;
e.g., Ramella et al. \cite{ram02}), a related topic is surely the
study of clustering properties in the large scale structure, usually
via the analysis of galaxy-galaxy correlation function.  A trend of
increasing clustering--strength with luminosity, independently of
morphological--type segregation, was found by Iovino et
al. (\cite{iov93}); see also, e.g., Loveday et al. (\cite{lov95}). In
particular, also for NOG galaxies, luminosity segregation is found
independently for both early-- and late--type galaxies (Giuricin et
al. \cite{giu01}).  Very recently, those results have been strongly
strengthened by the detailed analysis of the 2dF Galaxy Redshift
Survey, based on spectral types (Norberg et al. \cite{nor02}). Norberg
et al. find that the clustering strength increases with luminosity in
a similar way in both early and late types; however late types are
poorly represented at very high luminosities ($\sim 4 L^*$, see their
Figs.~9 and 10), where the clustering strength is the strongest. Thus
their results are not in contradiction with our finding of a
particularly strong luminosity segregation in very early--type
galaxies, since many of these galaxies are really very luminous and so
are expected to be highly clustered.

Most studies about luminosity segregation in galaxy systems or in the
large scale structure are based on visible light wavelengths. Near-IR
magnitudes are less sensitive to the effects of increased SFR in the
past, and thus are a good tracer of the galaxy stellar-mass (e.g.,
Gavazzi et al. \cite{gav96}) and result particularly useful in the
study of the evolution of clustering separately from that of the SFR.
Very interestingly, we also find evidence of luminosity segregation in
the K band, quantitatively similar to that in the B band. Beyond the
effect for very early--type galaxies, Fig.~\ref{fig7} shows that also
middle/late--type spirals have a strong segregation in the
magnitude--sample ($\Delta M_{\rm K}=-0.29\pm0.08$ for $2.5\le
T<5.5$), but the analysis of the volume--limited sample does not
confirm this result.  

Our results in K band allow us to estimate the stellar--mass
segregation between group and field environment, which can be derived
from:

\begin{equation}
M_{\star,G}/M_{\star,F}=\frac{(M_{\star}/L_K)_G\cdot L_{K,G}}{(M_{\star}/L_K)_F\cdot L_{K,F}}.
\end{equation} 

\noindent It has been suggested that the K-band
stellar-mass--to--light ratio can still vary by as much as a factor of
two over a range of galaxy Hubble type, color, and star formation
histories (e.g., Madau et al. \cite{mad98}), and gives values of about
unity for different IMF (e.g., Cole et al. \cite{col01}).  To estimate
the fraction of stellar--mass segregation due to the luminosity only,
i.e. independent of morphology--segregation, we adopt:
$(M_{\star}/L_K)_G=(M_{\star}/L_K)_F$ and $\Delta M_{\rm K}=-0.12$ mag
(i.e. the value found for the GROUP--ST-S-FIELD segregation) and
obtain $(M_{\star,G}-M_{\star,F})/\overline{M_{\star}} \sim 10\%$.  To
estimate the whole stellar--mass segregation we must consider $\Delta
M_{\rm K}=-0.35$ mag (i.e., the global luminosity segregation obtained
from the GROUP--S-FIELD comparison) and the fact that
$(M_{\star}/L_K)_G\ne(M_{\star}/L_K)_F$ due to the different
morphological content.  Assuming the values $M_{\star}/L_K$ observed
in nearby galaxies of early to late morphological--types by Charlot
\cite{cha96} (see Fig.~2 of Madau et al. \cite{mad98}), we obtain
$(M_{\star}/L_K)_G/(M_{\star}/L_K)_F \sim 1.2$.  We obtain a global
stellar--mass segregation of $\sim 50\%$. Note that this result is
independent of the absolute values of the assumed $M_{\star}/L_K$, but
depends on the variation of $M_{\star}/L_K$ with morphological type:
as for values we assume above, there is a variation of a factor $\sim
1.5$ from late to early types. Assuming a factor of $\sim 2$ (see
Fig.~1 of Bell \& de Jong \cite{bel01}, right--down panel) we obtain a
value of global stellar--mass segregation of $\sim 60\%$.

Colors are good tracers of the variations in the stellar populations
(e.g., Kennicutt \cite{ken83}).  To compare our results to those
recently obtained for environmental effects on colors or on SFR, we
considered that NOG groups have very low mean density.  Using the
values of member number and size we can estimate a projected density
of $\sim 5-6$ galaxies h$^2$ Mpc$^{-2}$ in groups of both the
magnitude-- and volume--limited samples.  This value is only slightly
higher than the characteristic density beyond which there is no
environmental effect in SFR, see our estimate of $\sim 3$ h$_{70}$
Mpc$^{-2}$ with the value of 1 galaxy h$_{70}$ Mpc$^{-2}$ for galaxies
with $M_{\rm B}<-19+5$\,log\,h$^2_{70}$ (Lewis et al. \cite{lew02};
see also G\'omez et al. \cite{gom03}). The value of the mean density
of our groups corresponds thus to the density of clusters at about 1-2
virial radii (see Fig.~7 of G\'omez et al. \cite{gom03}).

We find that group galaxies are redder than field galaxies, 
$\Delta$(B--K)$\sim 0.4$ mag. The comparison with B--R gradients
found by Pimbblet et al. (\cite{pim02}) is not straightford, but the
effect we find seems to be larger.  In fact, Pimbblet et al. find no,
or small, color difference between cluster regions at 2-3
h$_{50}^{-1}$ Mpc (whose density is comparable to that of our groups)
and less dense regions, $\Delta$(B--R)$\lesssim 0.05$ mag
(see their Table~3).

However, the color difference we measure seems to be largely induced
by the morphological segregation. When taking into account the
morphological segregation we find a difference of only
$\Delta$(B--K)=0.07 and $\Delta$(B--K)=0.21 mag in the magnitude-- and
volume--limited samples, respectively, and this difference is only
poorly significant (at $< 85\%$ according to the KS-test;
see Sect.~5).  Table~\ref{tab2} shows that only for very early--type
galaxies there is a 2sigma difference in both samples,
$\Delta$(B--K)=$0.15\pm 0.07$ mag.  Thus, we find very poor evidence
that color segregation is independent of morphological
segregation. Pimbblet et al. (\cite{pim02}) suggest that the
morphological segregation cannot induce the whole color segregation,
but their result concerns, in particular, the densest environments
(see their Tables~3 and 5).  Similarly, that the SFR--density relation
is not exclusively a result of the morphology--density relation is
suggested by the comparison between field and cluster internal
regions (within 1-2 virial radii; Balogh et al.  \cite{bal98}; G\'omez
et al. \cite{gom03}), a comparison to date based on a rather
crude morphological binning.  Thus, these results are not
incompatible with what we find analyzing less dense environments.
In the case the above differences between environments of different
density will be confirmed by future studies, one might speculate that
color segregation of galaxies is the sum of two effects, one which
depends on the morphology segregation and characterizes a large range
of environmental densities, from cluster cores to very poor groups,
and an additional effect, independent of morphological segregation,
which is strictly connected to cluster and/or very dense regions.  The
latter effect could not be appreciated in low density
environments.

\section{Summary}

We compare galaxy properties in group and field environments both in a
magnitude--limited sample (B$<14$ mag; $2000<cz<6000$ \ks; 959 vs. 2035
galaxies) and a volume--limited sample ($M_{\rm B} < -19.01+5$\,log\,h
mag; $2000<cz<4000$ \ks; 369 vs. 548 galaxies) extracted from the NOG
catalog (Giuricin et al. \cite{giu00}). For all these galaxies, blue
corrected magnitudes and morphological types are available. We
cross-correlate the NOG catalog with the 2MASS second release thus
assigning K magnitudes to about a half of NOG galaxies and obtaining
B--K colors.

We analyze B and K luminosity segregation and color segregation in
relation with the morphological segregation.  In fact, groups contain
about two times more early--type galaxies (ellipticals, lenticulars,
early spirals) than field, and many fewer late--type galaxies (late
spirals and irregulars).

We summarize our main results.

\begin{itemize}

\item For both B and K bands, we find that group galaxies are, on
average, more luminous than field galaxies and this is 
not entirely a consequence of the morphological effect.  After
correcting for morphological segregation, the global luminosity
difference between group and field galaxies is only modest, about
$10\%$ in both B and K bands.

\item The luminosity segregation--effect is particularly strong for
very early--type galaxies: group and field galaxies differ for more
than $30\%$ in luminosity.

\item 
We find that group galaxies are redder than field galaxies, 
$\Delta$(B--K)$\sim 0.4$ mag.  This difference is largely induced by
the morphological effect. After correcting for the morphological
segregation, we find a smaller, poorly significant, difference (at a
c.l. of $\sim 80\%$).

\end{itemize}

We discuss our results considering that the analyzed groups represent
a very low density environment (mean density $\sim 5-6$ galaxies h$^2$
Mpc$^{-2}$), just above the threshold value below which, according to
recent studies, environmental effects do not act.  On the basis of
this and previous results, we speculate that color segregation of
galaxies might be the sum of two effects, one which depends on the
morphology segregation and characterizes a large range of
environmental densities, from cluster cores to very poor groups, and
an additional effect, independent of morphology, which is
intrinsically connected to cluster and/or very dense regions.

\begin{acknowledgements}
We thank Andrea Biviano and Cristina Chiappini for useful
discussions.  We also thank the anonymous referee for a careful
reading of the manuscript and his/her valuable suggestions.  Work
partially supported by the Italian Ministry of Education, University,
and Research (MIUR, grant COFIN2001028932 "Clusters and groups of
galaxies, the interplay of dark and baryonic matter"), and by the
Italian Space Agency (ASI). CM acknowledges financial support from the
Centre National de la Recherche Scientifique and Region PACA.  This
publication makes use of data products from the Two Micron All Sky
Survey, which is a joint project of the University of Massachusetts
and the Infrared Processing and Analysis Center/California Institute
of Technology, funded by the National Aeronautics and Space
Administration and the National Science Foundation.
\end{acknowledgements}

\end{document}